\documentclass{article}
\usepackage[a4paper]{geometry}
\usepackage{amsmath}

\newtheorem{theorem}{Theorem}

\newenvironment{proof}[1][Proof]{\noindent\textbf{#1.} }{\ \rule{0.5em}{0.5em}}
\input{tcilatex}
\begin{document}

\title{Adiabatic Theorem in the Case of Continuous Spectra}
\author{M. Maamache and Y. Saadi \\
%EndAName
\\
\textit{Laboratoire de Physique Quantique et Syst\`{e}mes Dynamiques,}\\
\textit{{Facult\'{e} des Sciences,Universit\'{e} Ferhat Abbas de S\'{e}tif},
S\'{e}tif 19000, Algeria}}
\date{}
\maketitle

\begin{abstract}
In this paper,we present a rigorous demonstration and discussion of the
quantum adiabatic theorem for systems having a non degenerate continuous
spectrum. A new strategy is initiated by defining a kind of gap, "\textit{a
virtual gap}", for the continuous spectrum through the notion of
eigendifferential (Weyl's packet) and using the differential projector
operator. Finally we obtain the validity condition of the adiabatic
approximation. \newline

\end{abstract}

The adiabatic theorem is one of the basic results in quantum theory \cite{1,
2}. It is concerned with quantum systems described by an explicitly, but
slowly, time-dependent Hamiltonian. There has been a sudden regain of
interest in the adiabatic theorem for itself among physicists when in 1984
M.\ V. Berry \cite{3}\ pointed out that if it was applied to Hamiltonians
satisfying $H\left( t_{1}\right) =H\left( t_{2}\right) $, it could generate
a phase factor having non trivial geometrical meaning. And more recently,
the adiabatic theorem has renewed its importance in the context of quantum
control \cite{4}, for example, concerning adiabatic passage between atomic
energy levels, as well as for adiabatic quantum computation \cite{5}.

There are several points of view for a discussion of the quantum adiabatic
theorem; each one offers interesting insight. As T. Kato \cite{2}\ has
pointed out, the contents of the adiabatic theorem embody two parts: first,
the existence of a virtual change of the system which may be called an
adiabatic transformation, and, second, the dynamical transformation of the
system goes over to an adiabatic transformation in the limit when the change
of the Hamiltonian is infinitely slow. The adiabatic theorem proof given by
M. Born and V. Fock \cite{1}, although very general, is still restricted by
the assumption of considering the purely discrete and non-degenerate
Hamiltonian's spectrum, except for accidental degeneracy caused by crossing.
These limitations are rather artificial from the physical point of view and
should be removed from Kato's derivation of the adiabatic theorem. Several
authors \cite{6}\ had formally extended the Kato's results on the
approximate validity of the adiabatic theorem when the time $T$, during
which the approximation takes place, is large but finite. G. Nenciu \cite{9}
demonstrated the adiabatic theorem for bounded Hamiltonians. Later, J. E.
Avron and A. Elgart \cite{10} showed that the adiabatic theorem holds for
unbounded Hamiltonians as well and applied it to deal with the quantum Hall
effect.

Let us simply recall here that the works following that of Born and Fock 
\cite{1} by Kato \cite{2}, Garrido \cite{6}, Nenciu \cite{9} and J. E. Avron
et al. \cite{9} have led to a formulation of the adiabatic theorem under the
usual gap assumption $g_{nm}\left( t\right) =E_{n}\left( t\right)
-E_{m}\left( t\right) $, between level $n$ and $m$. One may then state that
a general validity condition for adiabatic behavior is well controlled as
follows: the larger is the quantity $\underset{0\leq t\leq T,\ m}{\min }%
\left[ g_{nm}\left( t\right) \right] $ the smaller will be the transition
probability.

Despite the existence of extensive literature on rigorous proofs of
estimates needed to justify the adiabatic approximation \cite{2, 9, 10, 12},
doubts have been raised about its validity \cite{14} leading to confusion
about the precise condition needed to use it \cite{15}. In part, this is
because some papers emphasize different aspects, such as the asymptotic
expansion, the replacement of the requirement of non-degenerate ground state
by a spectral projection separated from the rest of the spectrum, dependence
of first order estimates on the spectral gap, and even extensions to systems
without a gap. Adiabatic theorem without gap conditions is know to be true 
\cite{10}, however, in general, no estimates on the error terms are
available. J. E. Avron and A. Elgart have shown in ref.\cite{10}\ that the
adiabatic theorem holds provided the spectral projection is of finite rank
independently of any spectral considerations. A similar result was proven by
F. Bornemann \cite{18} for discrete Hamiltonian when the set of eigenvalues
crossings is of measure zero\ in time. The limitation of these approaches is
that, in general, no estimate can be made on the rate at which the adiabatic
regime is attained \cite{10}. The gap condition is generally associated to
spectral stabilities\textbf{.} Consequently, the situation where the gap
does not exist will led to spectral instabilities. Thus it is difficult to
establish smooth spectral projections which is a necessary condition for the
validity of the adiabatic theorem in the practical applications. In fact,
the generalized adiabatic theorem, according to J. E. Avron and A. Elgart's
approach \cite{10}, is much more appropriate for the systems without a gap
condition and which have a discrete origin.

In this letter, we present a straightforward, yet rigorous, proof of the
adiabatic theorem and adiabatic approximation for systems whose Hamiltonian
has a completely continuous spectrum supposed non-degenerated for reasons of
simplicity and which checks a certain number of conditions which will be
given later on.

In the case of continuous spectrum we cannot numerate eigenvalues and
eigenfunctions, they are characterised by the value of the physical quantity
in the corresponding state. Althoug the eigenfunctions $\varphi \left(
k;t\right) $ of the operators with continuous spectra cannot be normalised
in the usual manner as is done for the functions of discrete spectra, one
can construct with the $\varphi \left( k;t\right) $ new quantities -
theWeyl's\ \textit{eigendifferentials (wave packets)- }\cite{19} which
possess the properties of the eigenfunction of discrete spectrum. The
eigendifferentials are defined by the equation%
\begin{equation}
\left\vert \delta \varphi \left( k;t\right) \right\rangle =\overset{k+\delta
k}{\underset{k}{\int }}\left\vert \varphi \left( k^{\prime };t\right)
\right\rangle dk^{\prime }.  \label{eigendiff}
\end{equation}%
They divide up the continuous spectrum of the eigenvalues into finite but
sufficiently small discrete regions of size $\delta k$ (see Fig.\ref%
{Decomposition}).\FRAME{ftbphFUX}{0.8769in}{1.785in}{0pt}{\Qcb{\textbf{%
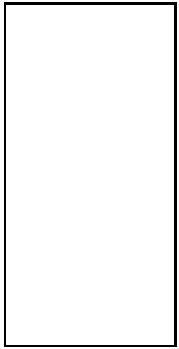
Decomposition of the continuous spectrum. }}}{\Qlb{Decomposition}}{Figure}{%
\special{language "Scientific Word";type "GRAPHIC";maintain-aspect-ratio
TRUE;display "USEDEF";valid_file "T";width 0.8769in;height 1.785in;depth
0pt;original-width 0.8441in;original-height 1.7478in;cropleft "0";croptop
"1";cropright "1";cropbottom "0";tempfilename
'JZ5FR500.wmf';tempfile-properties "XPR";}}

The eigendifferential (\ref{eigendiff}) is a special wave packet which has
only a finite extension in space; hence, it vanishes at infinity and
therefore can be seen in analogy to bound states. Furthermore, because the $%
\delta \varphi $ have finite spatial extension, they can be normalized. Then
in the limit $\delta k\rightarrow 0$,\ a meaningful normalization of the
function $\varphi $ themselves follows: the normalization on $\delta $
functions.

For $\delta k$, a small connected range of value of the parameter $k$ (this
corresponds to a group of "neighboring" states, see Fig.\ref{Decomposition}%
), the operator%
\begin{equation}
\delta P\left( k;t\right) =\underset{k}{\overset{k+\delta k}{\dint }}%
\left\vert \varphi \left( k^{\prime };t\right) \right\rangle \left\langle
\varphi \left( k^{\prime };t\right) \right\vert dk^{\prime }
\label{projector}
\end{equation}%
represents the projector (the differential projection operator \cite{20})
onto those states contained in the interval and characterized by the values
of the parameter $k$\ within the range of values $\delta k$. The action of $%
\delta P\left( k;t\right) $ on a wavefunction $\left\vert \psi \left(
t\right) \right\rangle $ is defined by

\bigskip

\bigskip 
\begin{equation}
\delta P\left( k;t\right) \left\vert \psi \left( t\right) \right\rangle =%
\underset{k}{\overset{k+\delta k}{\dint }}C\left( k^{\prime };t\right)
\left\vert \varphi \left( k^{\prime };t\right) \right\rangle dk^{\prime }.
\end{equation}

The application of the differential projection operator $\delta P\left(
k;t\right) $ causes thus the projection of the wavefunction onto the domain
of states $\varphi \left( k;t\right) $ which is characterized by $k$\ values
within the $\delta k$ interval .Before proceeding further, we give the
statement of the adiabatic theorem.

Let us call $U_{T}\left( s\right) $ the evolution operator where $s$ is the
fictitious time and $T$ is the time interval during which the evolution of
the system takes place%
\begin{equation}
i\hbar \frac{\partial }{\partial s}U_{T}\left( s\right) =TH\left( s\right)
U_{T}\left( s\right) ,  \label{OperEvolution}
\end{equation}%
and the slowly time-dependent Hamiltonian $H\left( s\right) =\int E\left(
k,s\right) \left\vert \varphi \left( k,s\right) \right\rangle \left\langle
\varphi \left( k,s\right) \right\vert dk\ $, $0\leq s\leq 1$, has a purely
continuous spectrum $E\left( k,s\right) $.

If the following conditions are fulfilled

\begin{description}
\item[(i)] As it is mentioned earlier (see Fig.\ref{Decomposition}) the
continuous spectrum is divided into discrete regions of size $\delta k$, we
must define or create a gap of energy for the continuous spectrum, in other
words, the size $\delta k$ is chosen so that%
\begin{equation}
E\left( k;s\right) -E\left( k^{\prime };s\right) >>\frac{1}{T}\ ,\ \forall
k^{\prime }\notin \left[ k,k+\delta k\right] .  \label{GapWidth}
\end{equation}

\item[(ii)] We assume that the eigenvalues are piecewise differentiable in
the parameter $s$, and there is no level crossing throughout the transition
(see Fig.\ref{evolution}), in other words:%
\begin{equation}
E\left( k^{\prime };s\right) \neq E\left( k^{\prime \prime };s\right) \ \
/s\in \left[ 0,1\right] ,k^{\prime }\in \left[ k,k+\delta k\right] ,\
k"\notin \left[ k,k+\delta k\right] .  \label{Noncrossing}
\end{equation}
\end{description}

\FRAME{ftbpFU}{3.0779in}{2.0358in}{0pt}{\Qcb{Evolution of a range of energy
of width $\protect\delta k$ as a function of time. }}{\Qlb{evolution}}{Figure%
}{\special{language "Scientific Word";type "GRAPHIC";maintain-aspect-ratio
TRUE;display "USEDEF";valid_file "T";width 3.0779in;height 2.0358in;depth
0pt;original-width 3.0338in;original-height 1.9969in;cropleft "0";croptop
"1";cropright "1";cropbottom "0";tempfilename
'JZ5FR501.wmf';tempfile-properties "XPR";}}

\begin{description}
\item[(iii)] The derivatives $\frac{\partial }{\partial s}\delta P\left(
k;s\right) $ and $\frac{\partial ^{2}}{\partial s^{2}}\delta P\left(
k;s\right) $ are well defined and continuous in the interval $0\leq s\leq 1$.
\end{description}

Under these conditions it is possible to prove the adiabatic theorem:

\begin{theorem}
If the quantum system with time-dependent Hamiltonian having a non
degenerate continuous spectrum is initially in an eigenstate $\left\vert
\varphi \left( k,0\right) \right\rangle $ of $H\left( 0\right) $ and if $%
H\left( s\right) $ evolves slowly enough then the state of the system at any
time $s$ will remain in the interval $\left[ k,k+\delta k\right] $.
\end{theorem}

The adiabatic theorem can be formally written, at the first order, in terms
of the evolution operator as%
\begin{equation}
\forall k:\underset{T\rightarrow \infty }{\ \lim }U\left( s\right) \delta
P\left( k;0\right) =\delta P\left( k;s\right) \underset{T\rightarrow \infty }%
{\lim }U\left( s\right) +O\left( \frac{1}{T}\right) .  \label{Theorem}
\end{equation}

Notice that if, initially, the system is in the state $\left\vert \varphi
\left( k,0\right) \right\rangle $ so that $H\left( 0\right) \left\vert
\varphi \left( k,0\right) \right\rangle =E\left( k,0\right) \left\vert
\varphi \left( k,0\right) \right\rangle $ and expanding an arbitrary state
vector on the basis of the instantaneous quasi-eigenfunction, then (\ref%
{Theorem}) implies%
\begin{equation}
\underset{T\rightarrow \infty }{\ \lim }U\left( s\right) \left\vert \varphi
\left( k,0\right) \right\rangle =\delta P\left( k;s\right) \underset{%
T\rightarrow \infty }{\lim }U\left( s\right) \left\vert \varphi \left(
k;0\right) \right\rangle
\end{equation}%
and in the limit $T\rightarrow \infty $ the state $U\left( s\right)
\left\vert \varphi \left( k;0\right) \right\rangle =\underset{k}{\overset{%
k+\delta k}{\dint }}C\left( k^{\prime };s\right) \left\vert \varphi \left(
k^{\prime };s\right) \right\rangle dk^{\prime },$ belongs to the subspace
generated by the states $\left\vert \varphi \left( k;s\right) \right\rangle $
pertaining to the interval $\left[ k,k+\delta k\right] .$

\begin{proof}
The demonstration that we present follows the same approach developed in ref.%
\cite{20} for the discrete case. To this effect, we introduce a unitary
operator $A\left( s\right) $ having the property%
\begin{equation}
\delta P\left( k,s\right) =A\left( s\right) \delta P\left( k,0\right)
A^{+}\left( s\right) \qquad \forall k\in \Re .  \label{Rotating}
\end{equation}%
It is completely defined by the initial condition $A\left( 0\right) =I$ and
the differential equation%
\begin{equation}
i\hbar \frac{\partial }{\partial s}A\left( s\right) =K\left( s\right)
A\left( s\right) .  \label{EqGene}
\end{equation}%
The operator $K\left( s\right) $ obeys the following commutation relation:%
\begin{equation}
i\hbar \frac{\partial }{\partial s}\delta P\left( k,s\right) =\left[ K\left(
s\right) ,\delta P\left( k,s\right) \right] ,  \label{Commut}
\end{equation}%
and is determined \ without ambiguity if we add the following supplementary
condition:%
\begin{equation}
\left\langle \varphi \left( k;t\right) \left\vert K\left( t\right)
\right\vert \varphi \left( k^{\prime };t\right) \right\rangle =0,\ \forall
k^{\prime }\in \left[ k,k+\delta k\right] ,  \label{SuppCond}
\end{equation}%
equation that yields the following expression%
\begin{equation}
K\left( t\right) =i\hbar \int \left[ 1-\delta P\left( k;t\right) \right]
\left\vert \dot{\varphi}\left( k;t\right) \right\rangle \left\langle \varphi
\left( k;t\right) \right\vert dk.  \label{Generator}
\end{equation}%
The unitary transformation $A^{+}\left( s\right) $, applied to the operators
and the vectors of the Schr\"{o}dinger's picture, produces a new picture:
the rotating axis picture:%
\begin{equation}
H^{\left( A\right) }\left( s\right) =A^{+}\left( s\right) H\left( s\right)
A\left( s\right) =\int E\left( k,s\right) \left\vert \varphi \left(
k,0\right) \right\rangle \left\langle \varphi \left( k,0\right) \right\vert
dk,  \label{HamiltA}
\end{equation}%
similarly $K^{\left( A\right) }\left( s\right) $ becomes 
\begin{equation}
K^{\left( A\right) }\left( s\right) =A^{+}\left( s\right) K\left( s\right)
A\left( s\right) .  \label{GeneratA}
\end{equation}%
The evolution operator in this new "representation" is $U^{\left( A\right)
}\left( s\right) =A\left( s\right) U_{T}\left( s\right) .$It is defined by 
\begin{equation}
i\hbar \frac{\partial }{\partial s}U^{\left( A\right) }\left( s\right) =%
\left[ TH^{\left( A\right) }\left( s\right) -K^{\left( A\right) }\left(
s\right) \right] U^{\left( A\right) }\left( s\right) ,\text{ \ \ \ \ \ \ \ }%
U^{\left( A\right) }\left( 0\right) =I.  \label{OperEvolutionA}
\end{equation}%
Since $H^{\left( A\right) }\left( s\right) $ and $K^{\left( A\right) }\left(
s\right) $ are $T$-independent, it is to be expected that in the $%
T\rightarrow \infty $ limit the first term of the right hand side in (\ref%
{OperEvolutionA}) dominates. We can approximately solve such an equation in
which we go over to a time-dependent reference frame following the axis
which diagonalize $H^{\left( A\right) }\left( s\right) $. We define $\Phi
_{T}\left( s\right) $ via%
\begin{equation}
i\hbar \frac{\partial }{\partial s}\Phi _{T}\left( s\right) =TH^{\left(
A\right) }\left( s\right) \Phi _{T}\left( s\right) ,  \label{EquPhi}
\end{equation}%
in which the solution may be written, with the initial condition $\Phi
_{T}\left( 0\right) =I$, as%
\begin{equation}
\Phi _{T}\left( s\right) =\int \exp \left[ -\frac{iT\alpha \left( k,s\right) 
}{\hbar }\right] \left\vert \varphi \left( k,0\right) \right\rangle
\left\langle \varphi \left( k,0\right) \right\vert dk,  \label{Phi}
\end{equation}%
where $\alpha \left( k,s\right) =\int_{0}^{s}E\left( k,s^{\prime }\right)
ds^{\prime }.$ If, as we will see immediately, $U^{\left( A\right) }\left(
s\right) $ tends toward $\Phi _{T}\left( s\right) $ for large $T$, we will
have approximately%
\begin{equation}
U_{T}\left( s\right) \underset{T\rightarrow \infty }{\simeq }A\left(
s\right) \Phi _{T}\left( s\right) .  \label{AdiaTheo}
\end{equation}%
We go over to a second picture and we show that the remaining evolution
operator differs from the identity by terms $O\left( \frac{1}{T}\right) $.
Thus, we change to a last picture with operator$W\left( s\right) \equiv \Phi
_{T}^{+}\left( s\right) A^{+}\left( s\right) U_{T}\left( s\right) $, and $%
\left( -\right) $ generator $\bar{K}$ $\left( s\right) =\Phi _{T}^{+}\left(
s\right) A^{+}\left( s\right) K\left( s\right) A\left( s\right) \Phi
_{T}\left( s\right) $:%
\begin{equation}
i\hbar \frac{\partial }{\partial s}W\left( s\right) =\bar{K}\left( s\right)
W\left( s\right) ,\qquad W\left( 0\right) =I,  \label{Lastpicture}
\end{equation}%
equivalent to the integral equation%
\begin{equation}
W\left( s\right) =I+\frac{i}{\hbar }\int_{0}^{s}\bar{K}\left( s^{\prime
}\right) W\left( s^{\prime }\right) ds^{\prime }.  \label{IntegrEqu}
\end{equation}%
Now, we prove that in the limit $T\rightarrow \infty $ , $W\left( s\right)
=I+O\left( \frac{1}{T}\right) .$We begin by considering the operator $%
F\left( s\right) =\int_{0}^{s}\bar{K}\left( s^{\prime }\right) ds^{\prime }.$
Any operator (and in particular $F\left( s\right) $) admits the following
decomposition%
\begin{eqnarray}
F\left( s\right) &=&\int \int F\left( k,k^{\prime },s\right) dkdk^{\prime } 
\notag \\
&=&\int_{0}^{s}\int \int \left\langle \varphi \left( k,0\right) \left\vert 
\bar{K}\left( s^{\prime }\right) \right\vert \varphi \left( k^{\prime
},0\right) \right\rangle \left\vert \varphi \left( k,0\right) \right\rangle
\left\langle \varphi \left( k^{\prime },0\right) \right\vert dkdk^{\prime
}ds^{\prime }.  \label{Decompos}
\end{eqnarray}%
Using (\ref{Phi}) we obtain:%
\begin{equation}
F\left( k,k^{\prime },s\right) =\int_{0}^{s}\exp \left[ \frac{iT\left(
\alpha \left( k,s^{\prime }\right) -\alpha \left( k^{\prime },s^{\prime
}\right) \right) }{\hbar }\right] K^{\left( A\right) }\left( k,k^{\prime
}s^{\prime }\right) ds^{\prime }\qquad k^{\prime }\notin \left[ k,k+\delta k%
\right] ,  \label{Fdecompos}
\end{equation}%
an expression in which we have introduced the condition $k^{\prime }\notin %
\left[ k,k+\delta k\right] $ because, from (\ref{SuppCond}), we deduce $%
F\left( k,k^{\prime },s\right) =0$ for $k^{\prime }\in \left[ k,k+\delta k%
\right] .$

Let $k^{\prime }\notin \left[ k,k+\delta k\right] $, since $K^{\left(
A\right) }\left( k,k^{\prime }s^{\prime }\right) $ is a continuous function
of $s$, our assumption implies that $\alpha \left( k,s^{\prime }\right)
-\alpha \left( k^{\prime },s^{\prime }\right) $ is a continuous nonvanishing
monotonic function of $s$; after integrating (\ref{Fdecompos}) by parts we
obtain%
\begin{eqnarray}
F\left( k,k^{\prime },s\right) &=&\frac{\hbar }{iT}\left[ \left. \exp \left[ 
\frac{iT\left( \alpha \left( k,s^{\prime }\right) -\alpha \left( k^{\prime
},s^{\prime }\right) \right) }{\hbar }\right] \frac{K^{\left( A\right)
}\left( k,k^{\prime }s^{\prime }\right) }{E\left( k,s^{\prime }\right)
-E\left( k^{\prime },s^{\prime }\right) }\right\vert _{0}^{s}\right. - 
\notag \\
&&-\left. \int_{0}^{s}\exp \left[ \frac{iT\left( \alpha \left( k,s^{\prime
}\right) -\alpha \left( k^{\prime },s^{\prime }\right) \right) }{\hbar }%
\right] \frac{\frac{\partial }{\partial s^{\prime }}K^{\left( A\right)
}\left( k,k^{\prime }s^{\prime }\right) }{E\left( k,s^{\prime }\right)
-E\left( k^{\prime },s^{\prime }\right) }ds^{\prime }\right] ,
\label{PartIntegr}
\end{eqnarray}%
hence, according to the condition (\ref{GapWidth}),$\ F\left( k,k^{\prime
},s\right) $, $k^{\prime }\notin \left[ k,k+\delta k\right] $,
asymptotically converges toward $0$ as $\frac{1}{T}$. Summarizing, as $%
T\rightarrow \infty $ we have:%
\begin{equation}
F\left( s\right) =O\left( \frac{1}{T}\right) .  \label{Order}
\end{equation}%
Using (\ref{Lastpicture}), integration by parts turns (\ref{IntegrEqu}) into:%
\begin{equation}
W\left( s\right) =I+\frac{i}{\hbar }F\left( s\right) W\left( s\right) +\frac{%
1}{\hbar ^{2}}\int_{0}^{s}F\left( s^{\prime }\right) \bar{K}\left( s\right)
W\left( s^{\prime }\right) ds^{\prime },
\end{equation}%
since the last two terms in this equation contain the factor $F\left(
s\right) $, then for $T\rightarrow \infty $ and from $U_{T}\left( s\right)
=A\left( s\right) \Phi _{T}\left( s\right) W\left( s\right) $ we obtain 
\begin{equation}
U_{T}\left( s\right) \simeq A\left( s\right) \Phi _{T}\left( s\right) \left[
I+O\left( \frac{1}{T}\right) \right]  \label{AdiabaticTheo}
\end{equation}%
Finally (\ref{Phi}) implies $\Phi _{T}\left( s\right) \delta P\left(
k,0\right) =\delta P\left( k,0\right) \Phi _{T}\left( s\right) $ and hence $%
A\left( s\right) \Phi _{T}\left( s\right) \delta P\left( k,0\right) =A\left(
s\right) \delta P\left( k,0\right) \Phi _{T}\left( s\right) =\delta P\left(
k,s\right) A\left( s\right) \Phi _{T}\left( s\right) $. This concludes the
proof of the adiabatic theorem (\ref{Theorem}).
\end{proof}

\bigskip

If $T$ is sufficiently large, we can, \ in first approximation, replace $%
U\left( t_{1},t_{0}\right) $ by its asymptotic form:%
\begin{equation}
U\left( t_{1},t_{0}\right) =U_{T}\left( 1\right) \simeq A\left( 1\right)
\Phi _{T}\left( 1\right) .
\end{equation}%
This is called the adiabatic approximation. If the initial normalized state
is $\left\vert \varphi \left( k_{0},0\right) \right\rangle $, under this
approximation $U\left( t_{1},t_{0}\right) $ $\left\vert \varphi \left(
k_{0},0\right) \right\rangle \approx A\left( 1\right) \Phi _{T}\left(
1\right) \left\vert \varphi \left( k_{0},0\right) \right\rangle $. To
determine the validity of the adiabatic approximation for a given process,
we can estimate the error by computing the probability $\eta $ of finding
the system at time $t_{1}$ in a state different from $A\left( 1\right) \Phi
_{T}\left( 1\right) \left\vert \varphi \left( k_{0},0\right) \right\rangle $:%
\begin{equation}
\eta =\left\langle \varphi \left( k_{0},0\right) \left\vert U^{+}\left(
t_{1},t_{0}\right) A\left( 1\right) \Phi _{T}\left( 1\right)
Q_{0}A^{+}\left( 1\right) \Phi _{T}^{+}\left( 1\right) U\left(
t_{1},t_{0}\right) \right\vert \varphi \left( k_{0},0\right) \right\rangle ,
\label{Probabilty}
\end{equation}%
where $Q_{0}=I-\delta P\left( k_{0},0\right) $. This quantity may be
rewritten as%
\begin{equation}
\eta =\left\langle \varphi \left( k_{0},0\right) \left\vert W^{+}\left(
1\right) Q_{0}W\left( 1\right) \right\vert \varphi \left( k_{0},0\right)
\right\rangle .  \label{Probabilty1}
\end{equation}%
Solving (\ref{IntegrEqu}) iteratively and keeping only the first order term,
we find%
\begin{eqnarray}
\eta &\approx &\frac{1}{\hbar ^{2}}\left\langle \varphi \left(
k_{0},0\right) \left\vert F^{+}\left( 1\right) Q_{0}F\left( 1\right)
\right\vert \varphi \left( k_{0},0\right) \right\rangle  \notag \\
&=&\frac{1}{\hbar ^{2}}\int_{k\notin \left[ k_{0},k_{0}+\delta k_{0}\right]
}\left\vert \left\langle \varphi \left( k_{0},0\right) \left\vert F\left(
1\right) \right\vert \varphi \left( k,0\right) \right\rangle \right\vert
^{2}dk.  \label{Probabilty2}
\end{eqnarray}

Now, let us define a normalized time through the variable transformation $%
t=t_{0}+sT$ $\left( 0\leq s\leq 1\right) $, and the initial normalized state 
$\left\vert \varphi \left( k_{0},t_{0}\right) \right\rangle $ of $H\left(
t_{0}\right) $ with the eigenvalue $E\left( k_{0},t_{0}\right) $. Then,
using (\ref{Generator}), (\ref{Fdecompos}) and performing the change $%
s\rightarrow t$ in eq. (\ref{Probabilty2}) yields%
\begin{eqnarray}
\left\langle \varphi \left( k_{0},t_{0}\right) \left\vert F\left(
t_{1}\right) \right\vert \varphi \left( k,t_{0}\right) \right\rangle 
&=&i\hbar \int_{t_{0}}^{t_{1}}\exp \left\{ \frac{i}{\hbar }\int_{t_{0}}^{t}%
\left[ E\left( k_{0},t^{\prime }\right) -E\left( k,t^{\prime }\right) \right]
dt^{\prime }\right\}   \notag \\
&&\left[ \left\langle \varphi \left( k_{0},t\right) |\dot{\varphi}\left(
k,t\right) \right\rangle -\left\langle \varphi \left( k_{0},t\right)
\left\vert \delta P\left( k_{0},t\right) \right\vert \dot{\varphi}\left(
k,t\right) \right\rangle \right] dt.
\end{eqnarray}%
Since $\left\langle \varphi \left( k_{0},t\right) \left\vert \delta P\left(
k_{0},t\right) \right\vert \dot{\varphi}\left( k,t\right) \right\rangle =0$
for $k\notin \left[ k_{0},k_{0}+\delta k_{0}\right] $ equation (\ref%
{Probabilty2}) may be recast as%
\begin{equation}
\eta \approx \frac{1}{\hbar ^{2}}\int_{k\notin \left[ k_{0},k_{0}+\delta
k_{0}\right] }\left\vert i\hbar \int_{t_{0}}^{t_{1}}\exp \left\{ \frac{i}{%
\hbar }\int_{t_{0}}^{t}\left[ E\left( k_{0},t^{\prime }\right) -E\left(
k,t^{\prime }\right) \right] dt^{\prime }\right\} \left\langle \varphi
\left( k_{0},t\right) |\dot{\varphi}\left( k,t\right) \right\rangle
dt\right\vert ^{2}dk,
\end{equation}%
and the adiabatic approximation for $\left\vert \varphi \left(
k_{0},t_{0}\right) \right\rangle $ holds only if $\eta \ll 1$ which requires%
\begin{equation}
\delta \wp _{\left( k_{0}\rightarrow k,t\right) }=\left\vert i\hbar
\int_{t_{0}}^{t_{1}}\exp \left\{ \frac{i}{\hbar }\int_{t_{0}}^{t}\left[
E\left( k_{0},t^{\prime }\right) -E\left( k,t^{\prime }\right) \right]
dt^{\prime }\right\} \left\langle \varphi \left( k_{0},t\right) |\dot{\varphi%
}\left( k,t\right) \right\rangle dt\right\vert ^{2}\ll 1,\quad \forall
k\notin \left[ k_{0},k_{0}+\delta k_{0}\right] ,  \label{Transition}
\end{equation}%
the integral of eq. (\ref{Transition}) will be sufficiently small, if the
phase of the integrated function vibrates fast enough and the amplitude of
the integrated function is small enough, thus $\delta \wp _{\left(
k_{0}\rightarrow k,t\right) }$ is at the maximum 
\begin{equation}
\delta \wp _{\left( k_{0}\rightarrow k,t\right) }\approx \underset{t\in %
\left[ t_{0},t_{1}\right] }{\max }\left\vert \frac{\hbar \left\langle
\varphi \left( k_{0},t\right) |\dot{\varphi}\left( k,t\right) \right\rangle 
}{E\left( k_{0},t^{\prime }\right) -E\left( k,t^{\prime }\right) }%
\right\vert ^{2},\quad \forall k\notin \left[ k_{0},k_{0}+\delta k_{0}\right]
.  \label{max1}
\end{equation}

The condition $\eta \ll 1$ is therefore, in most cases, certainly satisfied
if $\underset{k\notin \left[ k_{0},k_{0}+\delta k_{0}\right] }{\max }\delta
\wp _{\left( k_{0}\rightarrow k,t\right) }\ll 1$ or equivalently,%
\begin{equation}
\underset{k\notin \left[ k_{0},k_{0}+\delta k_{0}\right] }{\max }\left\vert
\left\langle \varphi \left( k_{0},t\right) |\dot{\varphi}\left( k,t\right)
\right\rangle \right\vert \ll \underset{k\notin \left[ k_{0},k_{0}+\delta
k_{0}\right] }{\min }\left\vert E\left( k_{0},t^{\prime }\right) -E\left(
k,t^{\prime }\right) \right\vert ,\quad \forall t\in \left[ t_{0},t_{1}%
\right] ,  \label{AdiaApproximation}
\end{equation}%
with $\max $ and $\min $ taken over all $k\notin \left[ k_{0},k_{0}+\delta
k_{0}\right] $. Condition (\ref{AdiaApproximation}) may be taken as a
criterion for the validity of the adiabatic approximation in the case of a
continuous spectrum. This estimate of the adiabatic approximation could not
be made in the Avron-Elgart's approach \cite{10} as mentioned earlier.

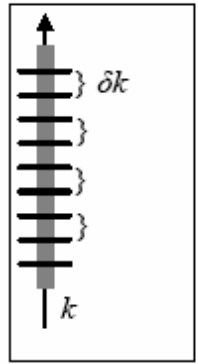
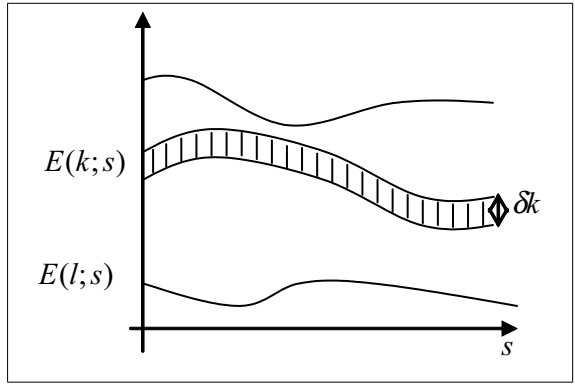
\end{document}